\documentclass[12pt]{article}

\usepackage[margin=1in]{geometry}
\usepackage{setspace}
\usepackage{times}
\usepackage{graphicx}
\usepackage{amsmath}
\usepackage{amssymb}
\usepackage{booktabs}
\usepackage{multirow}
\usepackage{float}
\usepackage{hyperref}
\usepackage[natbibapa]{apacite}
\usepackage{fancyhdr}
\usepackage{authblk}

\begin{document}

\thispagestyle{plain}

\begin{center}
{\Large\textbf{Visual Orientalism in the AI Era:\\From West-East Binaries to English-Language Centrism}}

\vspace{1.5em}

Zhilong Zhao$^{a,b,*}$, Yindi Liu$^{a,b}$\\

\vspace{1em}
{\small $^{a}$School of Journalism and Communication, South China University of Technology, Guangzhou, China}\\
{\small $^{b}$Guangdong--Hong Kong--Macao Greater Bay Area Research Institute of International Communication,}\\
{\small South China University of Technology, Guangzhou, China}\\

\vspace{0.8em}
{\small $^{*}$Corresponding author: Zhilong Zhao}

\vspace{1em}
{\small November 2025}
\end{center}

\vspace{1.5em}

\noindent\textbf{Abstract}

\noindent
Text-to-image AI models systematically encode geopolitical bias through visual representation. Drawing on Said's Orientalism and framing theory, we introduce \textit{Visual Orientalism}---the dual standard whereby AI depicts Western nations through political-modern symbols while portraying Eastern nations through cultural-traditional symbols. Analyzing 396 AI-generated images across 12 countries and 3 models, we reveal an evolution: Visual Orientalism has shifted from traditional West-versus-East binaries to \textit{English-language centrism}, where only English-speaking core countries (USA and UK) receive political representation while all other nations---including European powers---face cultural exoticization. This algorithmic reconfiguration of Orientalism operates through automated framing mechanisms shaped by the material conditions of AI development: English-language training data dominance and the concentration of AI development in English-speaking tech companies. Our findings demonstrate how AI systems function as agents of cultural representation that perpetuate and intensify historical power asymmetries. Addressing Visual Orientalism requires rethinking of algorithmic governance and the geopolitical structures embedded in AI training data.

\vspace{1em}


\newpage


\section{Introduction}

The rapid proliferation of text-to-image (T2I) AI models has democratized visual content creation, with platforms like DALL-E, Midjourney, and Stable Diffusion generating images at unprecedented scale \citep{bianchi2023easily}. As AI-generated imagery becomes embedded in social media, news, marketing, and education, these systems wield unprecedented power in shaping collective visual imagination \citep{luccioni2023stable}. However, emerging research reveals that these systems encode systematic biases in visual representation \citep{cho2023dall, naik2023social, zhou2024bias}. While existing studies have documented gender, racial, and occupational stereotypes, the \textit{cultural and geopolitical dimensions} of visual bias remain underexplored.

This study introduces the concept of \textit{Visual Orientalism} to examine how AI image generation models perpetuate dual standards in representing different countries. Drawing on framing theory \citep{entman1993framing, scheufele2007framing} and critical algorithm studies \citep{gillespie2014relevance, bucher2018if}, we conceptualize AI image generation systems as \textit{automated framing agents}---computational systems that encode framing decisions into algorithms operating at scale. These systems systematically select and organize visual symbols to construct particular narratives about nations and cultures. Building on Said's seminal work on Orientalism \citep{said1978orientalism} and recent scholarship on algorithmic orientalism \citep{kotliar2020data, crawford2021atlas}, we argue that AI systems systematically depict certain nations through \textit{political-modern-sovereign} visual symbols (e.g., national flags, government buildings), while portraying others through \textit{cultural-traditional-exotic} visual symbols (e.g., ethnic clothing, religious architecture). Critically, we investigate whether these boundaries align with traditional West-East binaries or have shifted in the AI era, potentially reflecting the concentration of AI development in English-speaking tech hubs. This dual standard operates through training data and model architectures that encode historical patterns of visual dominance, functioning as what \citet{just2017governance} term ``governance by algorithms''---the algorithmic construction of social reality through systematic selection and exclusion.

Through analysis of 396 AI-generated images across 12 countries, 11 concepts, and 3 T2I models, we demonstrate that Visual Orientalism operates through three mechanisms: (1) \textit{visual symbol selection bias}, (2) \textit{visual narrative framing}, and (3) \textit{visual power asymmetry}. Our study makes three contributions. First, addressing the gap between framing theory and algorithmic systems, we extend framing theory \citep{entman1993framing} to algorithmic visual production, demonstrating how AI systems function as automated framing agents perpetuating Orientalist frameworks. Second, responding to methodological challenges in analyzing AI-generated content at scale, we develop a multi-model AI-assisted content analysis combining four VLMs with entropy-based quality control. Third, we provide systematic evidence of Visual Orientalism across multiple dimensions, revealing gendered, spatial, and temporal patterns reflecting broader power asymmetries in global algorithmic infrastructure \citep{napoli2014automated}.

\section{Literature Review}

\subsection{From Textual to Visual Orientalism}

Edward Said's \textit{Orientalism} \citep{said1978orientalism} established how the West constructs the East as its cultural ``Other'' through binary oppositions justifying dominance. Said conceptualized Orientalism not as a static set of stereotypes but as a \textit{dynamic system of knowledge production} that maintains power asymmetries, with boundaries shifting as geopolitical structures evolve. Subsequent scholarship extended this framework to religious exoticization \citep{king2013orientalism}, multiple discourses \citep{varisco2011reading}, and the post-9/11 era \citep{dabashi2017post}.

Recent work extended Said's framework to computational systems. \citet{kotliar2020data} introduced the concept of \textit{data orientalism} to describe how algorithmic systems construct non-Western societies as fundamentally different through data collection and analysis practices. \citet{crawford2021atlas} demonstrated that AI training data naturalizes social hierarchies and reflects geopolitical power structures embedded in global media production. However, these extensions have not systematically examined the \textit{visual dimension} of algorithmic orientalism, nor investigated whether AI systems reproduce traditional orientalist boundaries or reconfigure them around new centers of power. Given the concentration of AI development in English-speaking tech companies and the dominance of English-language training data, Visual Orientalism boundaries may have shifted beyond traditional West-East binaries.

This visual dimension is particularly important because visual symbols carry distinct semiotic and political meanings. Visual representation is \textit{constitutive practice} that actively constructs meaning \citep{hall1997representation}, with visual symbols ``naturalizing'' cultural ideologies through denotative, connotative, and mythological layers \citep{barthes1977rhetoric, aiello2020visual}. We conceptualize Visual Orientalism as a \textit{continuum from political representation to cultural exoticization}, which enables us to capture gradations beyond the traditional West-East binary and identify unexpected patterns in how AI systems draw orientalist boundaries. This framework extends Said's theory to the visual-algorithmic domain, drawing on visual semiotics and symbolic politics to analyze how AI systems encode geopolitical hierarchies \citep{barthes1977rhetoric, aiello2020visual, kolsto2006national, butz2009national}.

\subsection{Algorithmic Framing and the Reproduction of Cultural Hierarchies}

Framing theory provides a foundational framework for understanding how AI systems construct Visual Orientalism. \citet{entman1993framing} defined framing as selecting and highlighting certain aspects of reality to promote particular interpretations, while \citet{scheufele2007framing} distinguished between frame building (how frames are constructed) and frame setting (how frames influence audiences). We conceptualize AI image generation systems as \textit{automated framing agents} that systematically select and organize visual symbols to construct particular narratives about nations and cultures, with framing decisions reflecting systematic patterns encoded in training data \citep{bianchi2023easily}.

Algorithmic framing differs fundamentally from human framing in its scale, consistency, and opacity. While human editors make framing decisions that may be influenced by unconscious biases, AI systems encode framing patterns through statistical regularities in training data \citep{schuhmann2022laion, birhane2021multimodal}, operating at massive scale with remarkable consistency \citep{bianchi2023easily, luccioni2023stable} and deliberate opacity \citep{gillespie2014relevance}. Once biases are encoded, this \textit{algorithmic consistency} means biased framing patterns are reproduced systematically across millions of images. More critically, algorithms function as powerful gatekeepers encoding particular ``ways of knowing'' that privilege certain information while marginalizing others \citep{bucher2018if}. In international communication, algorithmic systems operated by Western technology companies perpetuate cultural domination through this ``governance by algorithms''---the construction of social reality through systematic selection \citep{just2017governance}, naturalizing cultural hierarchies more effectively than explicit human editorial decisions \citep{napoli2014automated}.

Applied to Visual Orientalism, AI image generation systems function as \textit{automated framing agents}, systematically constructing certain nations as modern and sovereign while depicting others as traditional and exotic. These patterns emerge from training data encoding centuries of Orientalist visual production in Western media \citep{bianchi2023easily, birhane2021multimodal}. By learning statistical regularities from such biased data, AI systems internalize and reproduce them at unprecedented scale---what we term \textit{algorithmic Visual Orientalism}. Critically, the material conditions of AI development---English-language training data dominance, the disproportionate concentration of frontier AI labs and infrastructure in the United States, and the hegemony of English-speaking tech companies---may reconfigure Orientalism's boundaries around English-language centrism rather than traditional West-East binaries. This necessitates new forms of ``algorithmic accountability''---mechanisms for interrogating cultural assumptions encoded in computational systems \citep{diakopoulos2015algorithmic}.

\subsection{Visual Symbol Selection Bias in AI Image Generation}

Recent research documented systematic biases in text-to-image (T2I) models across gender, skintone, and geo-cultural dimensions \citep{wan2024survey, cho2023dall, naik2023social, bianchi2023easily}. These biases root in training data: LAION-400M shows Western overrepresentation while non-Western regions are underrepresented or stereotypically depicted \citep{birhane2021multimodal, schuhmann2022laion}, reflecting ``data orientalism''---the construction of non-Western societies as fundamentally different through algorithmic systems \citep{kotliar2020data}.

However, existing research has not systematically examined the \textit{political versus cultural symbol distinction} in AI-generated national representations. Political symbols---particularly national flags---serve as visual markers of sovereignty and political agency \citep{marshall2017flag, kolsto2006national}, while cultural symbols signify heritage and tradition but lack political authority \citep{cerulo1995identity}. The differential assignment of these symbol types constitutes \textit{symbolic power asymmetry}.

Symbolic representation operates across multiple dimensions. Flags vary not only in presence/absence but in \textit{visual prominence}, while cultural symbols interact with modernity levels. Existing research has relied on simple symbol counts, potentially missing these patterns. A multidimensional measurement approach weighting symbols by theoretical significance and visual prominence may reveal patterns invisible to simpler metrics.

Flags carry particular political significance as sovereignty markers, representing \textit{symbolic sovereignty}---the visual assertion of statehood that enables political agency \citep{marshall2017flag, kolsto2006national}. Critically, flags function as \textit{sovereignty signifiers} whose visual prominence signals hierarchical positioning within the international order: a prominently displayed flag asserts political legitimacy, while flag absence relegates nations to cultural curiosity rather than sovereign actors \citep{cerulo1995identity}.

If AI models systematically assign political symbols to certain countries while assigning cultural symbols to others, this constitutes \textit{visual symbol selection bias}---a form of algorithmic orientalism that creates \textit{symbolic power asymmetry} by granting some nations visual political legitimacy while relegating others to cultural exoticism. This bias operates through training data encoding historical patterns of visual dominance in global media. This leads to our first research question:

\textbf{RQ1}: To what extent do AI image generation models exhibit systematic differences in assigning political versus cultural symbols to different countries, and where do the boundaries of Visual Orientalism lie?

Given Said's (1978) emphasis on Orientalism as a dynamic system whose boundaries shift with geopolitical power structures, we adopt an exploratory approach to examine how Visual Orientalism manifests in AI-generated imagery. Rather than presupposing traditional West-East binaries, we investigate:

\begin{itemize}
    \item \textbf{RQ1a (Exploratory)}: How are political and cultural symbols distributed across different countries? Do traditional West-East distinctions hold, or have Visual Orientalism boundaries shifted in the AI era?
    \item \textbf{RQ1b (Exploratory)}: To what extent do different countries exhibit political sovereignty representation versus cultural exoticization, and what patterns emerge beyond the traditional West-East binary?
    \item \textbf{H1 (Directional)}: Countries will exhibit systematic differences in Visual Orientalism patterns, with effect sizes indicating meaningful symbolic power asymmetries in how AI systems represent national identities.
\end{itemize}

\subsection{Visual Narrative Framing and Intersectionality}

Beyond symbol selection, AI-generated imagery may perpetuate Visual Orientalism through \textit{narrative framing}---how visual elements are selected, organized, and contextualized to construct particular narratives. Visual framing theory \citep{hall1997representation, rodriguez2011levels, schill2012visual} examines how images construct meaning through multiple layers: denotative framing (what is shown), stylistic framing (how it is shown), and connotative framing (what cultural meanings are activated).

Applied to Visual Orientalism, narrative framing operates through systematic association of Western nations with modernity and contemporaneity, while Eastern nations are associated with tradition and historical continuity. This temporal framing reproduces what \citet{said1978orientalism} identified as a core feature of Orientalism: constructing the East as frozen in a pre-modern past while the West advances into the future. Visually, this manifests through depicting Western countries with contemporary architecture and civic institutions, while Eastern countries are depicted with historical monuments and traditional practices.

Visual Orientalism's effects are not uniform but intensify when intersecting with other social categories. Visual framing operates with particular intensity through gendered and temporal dimensions. T2I models systematically underrepresent women in professional contexts, reinforcing traditional gender roles \citep{gorska2023invisible}. This aligns with intersectionality theory \citep{crenshaw1989demarginalizing}, which argues that systems of oppression interact to produce unique forms of marginalization.

Applied to Visual Orientalism, intersectionality suggests certain categories---particularly gender and temporal contexts---are especially susceptible to orientalist framing. Eastern women may face dual marginalization, resulting in heightened cultural symbolization. Similarly, festivals provide opportunities for AI models to emphasize cultural difference, framing Western festivals as civic celebrations while depicting Eastern festivals as exotic spectacles.

This leads to our second research question:

\textbf{RQ2}: How does Visual Orientalism vary across different concepts (gender, time), and what intersectional patterns emerge?

\begin{itemize}
    \item \textbf{H2a}: Eastern women will be depicted with more cultural symbols and lower modernity levels than Western women (gendered Visual Orientalism).
    \item \textbf{H2b}: Eastern festivals will be depicted with more cultural symbols and lower modernity levels than Western festivals (temporal Visual Orientalism).
\end{itemize}

\subsection{Model-Level Variations in Text-to-Image Bias}

While prior research has documented systematic biases in text-to-image (T2I) models, an emerging body of evidence suggests that different T2I models exhibit varying levels and patterns of bias. \citet{luccioni2023stable} conducted a comparative analysis of multiple T2I models (Dall·E 2, Stable Diffusion v1.4 and v2), finding that while all models exhibit systematic biases, they differ in the extent to which they under-represent marginalized identities. \citet{bianchi2023easily} demonstrated that geographic and demographic biases persist even in models with explicit mitigation strategies, with stereotypical representations of non-Western regions appearing across different T2I systems. These findings suggest that Visual Orientalism is not a uniform phenomenon but may vary systematically depending on model-specific characteristics. Understanding model-level variations is crucial for determining whether Visual Orientalism is an inherent feature of current AI architectures or whether certain design choices can mitigate these biases.

Model-level variations are driven by differences in training data composition, model architecture, and development practices. \citet{luccioni2023stable} demonstrated that training data composition directly influences the types and magnitudes of biases exhibited by diffusion models, with models trained on different datasets or at different time periods exhibiting distinct bias patterns. \citet{friedrich2023fair} showed that instructing diffusion models on fairness during training can reduce demographic biases, but these interventions cannot eliminate bias entirely. \citet{zhou2024bias} found that even state-of-the-art generative AI models exhibit persistent biases across multiple dimensions, suggesting that current debiasing techniques have limited effectiveness. Critically, existing debiasing interventions focus primarily on demographic dimensions such as gender and race, while cultural and geopolitical dimensions remain largely unaddressed.

This raises a critical question with both theoretical and practical implications: if Visual Orientalism persists across models despite debiasing efforts, this would suggest that geopolitical biases are deeply embedded in training data and resistant to current mitigation strategies. Conversely, if certain models exhibit lower Visual Orientalism, this would provide evidence that design choices can reduce these biases. This leads to our third research question:

\textbf{RQ3}: Do different text-to-image models exhibit different levels of Visual Orientalism, and what factors explain these variations?

\begin{itemize}
    \item \textbf{H3a}: Different T2I models will exhibit significantly different levels of Visual Orientalism bias.
    \item \textbf{H3b}: Models with more recent training data and explicit debiasing strategies (e.g., GPT-Image-1) will exhibit lower Visual Orientalism bias than earlier models.
\end{itemize}

\section{Methodology}

To investigate whether AI-generated imagery reproduces Said's (1978) Orientalist framework, we designed a comparative study examining how three text-to-image models represent 12 countries across diverse contexts. Our approach balances two goals: capturing systematic representational bias while ensuring methodological rigor in analyzing visual content at scale. If Visual Orientalism operates algorithmically, we should observe consistent dual standards---Western nations depicted through political-modern-sovereign symbols, Eastern nations through cultural-traditional-exotic symbols---across models, countries, and concepts.

Our methodological strategy combines large-scale image generation (396 images) with multi-model AI-assisted content analysis. A key innovation is our use of four vision-language models as coders, with systematic quality assessment and expert validation. Drawing on recent advances in automated content analysis \citep{zhao2025automated, zhao2025hierarchical}, this approach enables rigorous examination of visual patterns while maintaining the interpretive validity essential for cultural analysis. This section proceeds in four parts: research design and data collection, coding procedures, quality assessment and validation, and analytical strategy.

\subsection{Research Design and Data Collection}

We constructed a balanced factorial design crossing three dimensions: \textit{countries} (12), \textit{concepts} (11), and \textit{T2I models} (3), yielding 396 unique images. Country selection balanced geographic diversity with theoretical relevance: 5 Western countries representing historical centers of Orientalist discourse: United States, United Kingdom, France, Germany (European colonial powers), and Australia (English-speaking settler-colonial nation), and 7 non-Western countries (China, India, Japan, South Korea, Brazil, Russia, Egypt) representing diverse regions subject to Orientalist discourse in Said's (1978) framework. Following Said's conceptualization of the ``Orient'' as a Western construction encompassing diverse non-Western regions, we use ``Eastern'' and ``non-Western'' interchangeably to refer to countries subject to Orientalist representation, while acknowledging the geographic imprecision of these terms. The 11 concepts span national identity (country itself), demographic groups (people, women, men, elderly, children, students), and cultural artifacts (cities, architecture, festivals, cuisine), enabling examination of Visual Orientalism across diverse representational contexts. We selected three T2I models representing different design philosophies: GPT-Image-1 (OpenAI), Midjourney, and NanoBanana. Image generation followed a standardized protocol using neutral prompts: ``[Concept] in [Country]'' (e.g., ``women in China''). All images were generated between September and October 2025 using default settings (1024$\times$1024 pixels).

\subsection{Coding Procedures}

We developed a structured coding scheme operationalizing Visual Orientalism across five dimensions, implemented through a multi-model VLM ensemble. We employed four VLMs to code each image: Qwen3-VL-32B-Instruct, GPT-5 (OpenAI), Gemini 2.5 Flash (Google), and Claude Haiku 4.5 (Anthropic). Multi-model ensemble coding reduces individual model biases and provides quality signals through cross-model agreement.

\textit{Coding Dimensions.} Our coding scheme operationalizes Visual Orientalism's core distinction: political-modern-sovereign representation of the West versus cultural-traditional-exotic representation of the East. Following expert validation that identified redundancy in an initial six-dimension scheme, we refined the coding to five dimensions: (1) Political Symbols (count); (2) Cultural Symbols (count); (3) Flag Appearance (0-4 scale); (4) Sovereignty Representation (binary); (5) Modernity Level (1-5 scale). Detailed definitions, coding rules, examples, and VLM prompts are provided in Supplemental Materials.

\textit{Composite Indices.} We constructed a \textit{Symbolization Index} (SI) capturing the balance between political and cultural representation:

\begin{equation}
SI = \frac{\text{Political Symbols} - \text{Cultural Symbols}}{\text{Political Symbols} + \text{Cultural Symbols} + 1}
\end{equation}

To more sensitively capture multidimensional Visual Orientalism, we constructed three weighted composite indices. The \textit{Political Sovereignty Index} (PSI) combines flag prominence (40\% weight), sovereignty representation (30\%), and political symbol count (30\%):

\begin{equation}
PSI = 0.4 \cdot \frac{\text{Flag Appearance}}{4} + 0.3 \cdot \text{Sovereignty} + 0.3 \cdot \frac{\text{Political Symbols}}{\max(\text{Political Symbols})}
\end{equation}

The \textit{Cultural Exoticization Index} (CEI) combines cultural symbol count (40\%), traditionality (30\%), and flag absence (30\%):

\begin{equation}
CEI = 0.4 \cdot \frac{\text{Cultural Symbols}}{\max(\text{Cultural Symbols})} + 0.3 \cdot \left(1 - \frac{\text{Modernity}}{5}\right) + 0.3 \cdot \left(1 - \frac{\text{Flag Appearance}}{4}\right)
\end{equation}

The \textit{Visual Orientalism Index} (VOI) combines these indices:

\begin{equation}
VOI = PSI - CEI
\end{equation}

\subsection{Quality Assessment and Validation}

To assess coding quality, we adopted a dual-signal framework combining model confidence and cross-model agreement \citep{zhao2025automated}. The four VLMs achieved high inter-model consistency (mean external entropy $H_{ext} = 0.35$, SD = 0.20). We implemented entropy-based prioritized sampling for expert validation, sampling 67 images (16.9\%) with priority given to high-disagreement cases (37 images with entropy $> 0.6$, 30 with $0.4 < $ entropy $ \leq 0.6$), concentrating validation resources on uncertain cases where AI models disagree.

Two expert coders independently coded the validation sample. Inter-coder reliability was excellent across all five dimensions (Krippendorff's $\alpha = 0.76$-$0.97$) \citep{krippendorff2018content}. AI-human agreement was substantial ($\alpha = 0.82$), with multi-model consensus outperforming any single VLM. We applied the Hierarchical Error Correction (HEC) framework \citep{zhao2025hierarchical} to iteratively refine coding prompts. Sensitivity analyses comparing results using AI codes (n=396) versus expert codes (n=67) showed substantively identical effect sizes (mean difference in Cohen's d = 0.03), confirming reliable coding quality. For the final analysis, we used multi-model consensus codes for all 396 images.

\subsection{Analytical Strategy}

Data analysis proceeded in three stages corresponding to our research questions. First, we conducted descriptive analysis of symbol distribution by country and concept to address RQ1. Second, for RQ1's primary hypothesis tests, we used independent-samples t-tests for East-West comparisons at the country level (aggregating each country's 33 images into country means, df=10), with effect sizes reported using Cohen's $d$. We used weighted composite indices (PSI, CEI, VOI) as primary outcome measures. For analyses involving specific concepts (e.g., festivals, gender), we conducted image-level analyses to preserve within-concept variation. Third, we examined intersectionality patterns using two-way ANOVA with region (East/West) and gender (women/men) as factors at the image level (RQ2), and model-level variations using mixed-effects ANOVA with model as a fixed factor and country as a random factor at the image level, followed by post-hoc Tukey HSD tests (RQ3). For post-hoc tests involving multiple pairwise comparisons, we used Tukey's HSD method to control family-wise error rate. Statistical significance was assessed at $\alpha = 0.05$.

\section{Results}

We present results in three sections corresponding to our research questions: visual symbol selection bias (RQ1), visual narrative framing and intersectionality (RQ2), and model-level variations (RQ3). All tests use $\alpha = 0.05$.

\subsection{Visual Symbol Selection Bias (RQ1)}

\subsubsection{Political versus Cultural Symbol Distribution}

Our exploratory analysis (RQ1a) revealed systematic differences in how AI systems assign political versus cultural symbols to different countries. Western countries averaged 0.76 political symbols per image (SD = 0.70) compared to Eastern countries' 0.20 (SD = 0.16), showing a large effect size ($t(10) = 2.07$, $p = 0.065$, Cohen's $d = 1.10$). Conversely, Eastern countries averaged 4.02 cultural symbols per image (SD = 0.59) compared to Western countries' 2.87 (SD = 0.86), showing a large effect size that reached statistical significance ($t(10) = -2.76$, $p = 0.020$, $d = 1.56$). These differences indicate meaningful symbolic power asymmetries, supporting H1.

\begin{figure}[H]
\centering
\includegraphics[width=\textwidth]{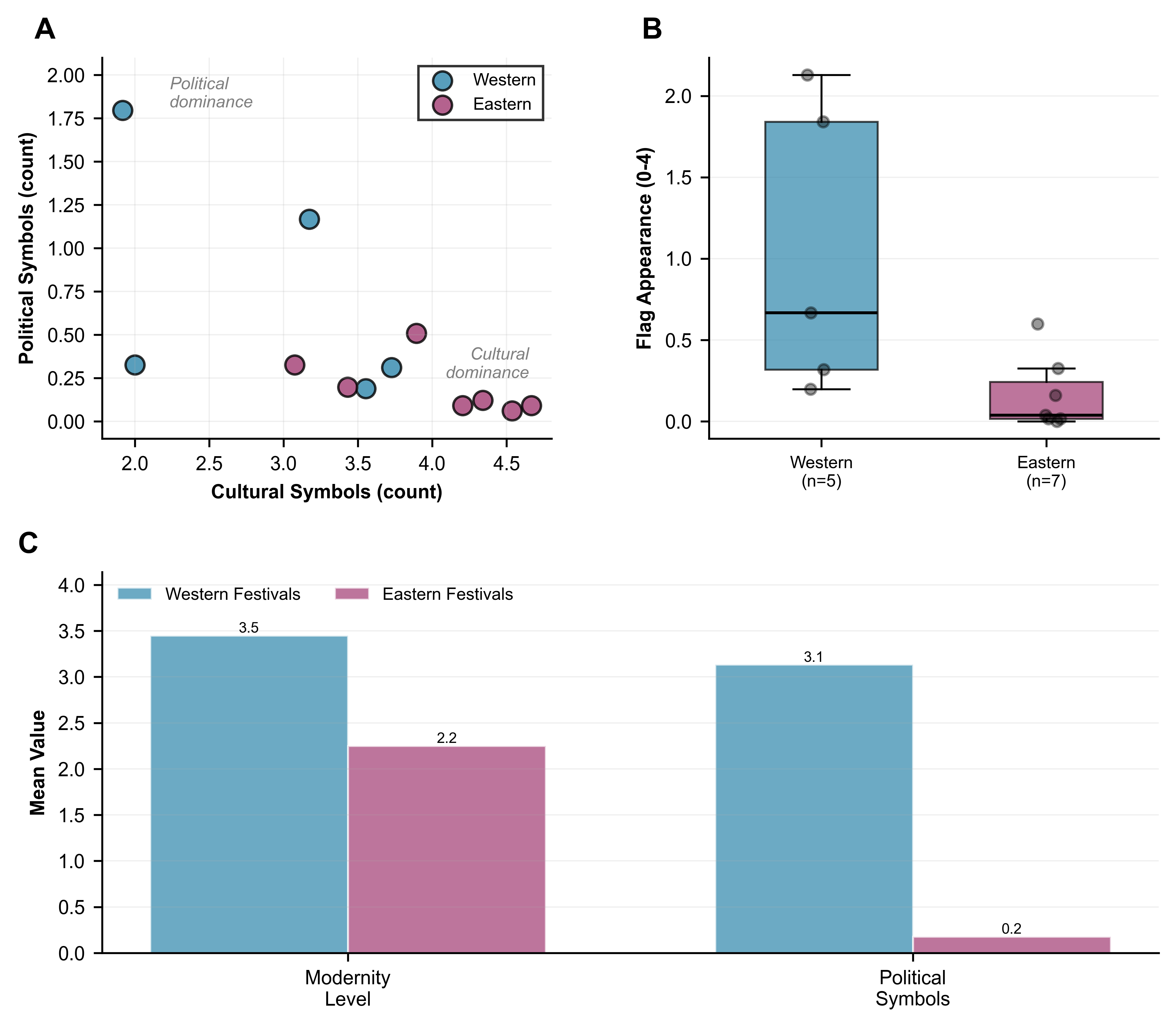}
\caption{Main Findings of Visual Orientalism in AI Image Generation. (A) Symbol distribution: Western countries cluster in the lower-left region with higher political representation relative to Eastern countries, while Eastern countries concentrate in the lower-right region showing strong cultural dominance with minimal political representation. (B) Flag effect: Western countries M=1.03 vs. Eastern M=0.16, $d=1.33$, $p=0.031$. (C) Festival effect: Western festivals are framed as more modern and political, while Eastern festivals are depicted as more traditional and cultural, with a large modernity gap between regions ($d=1.23$, $p<0.001$). *$p<0.05$, **$p<0.01$, ***$p<0.001$.}
\label{fig:1}
\end{figure}

However, Figure~\ref{fig:1}A reveals a pattern that challenges traditional West-East binaries: while Eastern countries cluster in the lower-right quadrant (high cultural symbols, low political symbols), Western countries show substantial heterogeneity. The United States and United Kingdom appear in the upper-left region (political symbol dominance), France and Germany cluster near Eastern countries in the lower-right region (cultural symbol dominance), while Australia occupies the lower-left quadrant (low cultural, low political). This heterogeneity within the ``West'' suggests that Visual Orientalism boundaries have shifted beyond traditional geographic categories.

\subsubsection{Flag Appearance as Sovereignty Marker}

National flags---the quintessential symbol of political sovereignty---appeared with dramatically different frequencies. Western countries averaged 1.03 flag appearances per image (SD = 0.89) versus Eastern countries' 0.16 (SD = 0.22), showing a large effect size that reached statistical significance ($t(10) = 2.50$, $p = 0.031$, Cohen's $d = 1.33$). As shown in Figure~\ref{fig:1}B, Western images incorporated flags substantially more frequently than Eastern images, with the USA showing the highest rate (M = 2.13) and several Eastern countries (China, Japan, India) showing near-zero rates (M < 0.10).

However, this pattern again reveals Western heterogeneity: the United States (M = 2.13) and United Kingdom (M = 1.83) show high flag appearance rates, while France (M = 0.31) and Germany (M = 0.19) show rates comparable to Eastern countries. This suggests that flag-based sovereignty representation is concentrated in English-speaking core countries rather than distributed across ``the West'' as a whole.

Flag appearance varied systematically by concept. For ``country itself,'' flags appeared prominently in English-speaking core countries' images but minimally in other countries' images. For demographic concepts, English-speaking core countries' images showed flags while other countries' images rarely did. This pattern suggests AI models associate English-speaking core nations with political sovereignty across all contexts, while other nations receive political representation only selectively (see Supplemental Materials Table S1).

\subsubsection{Country-Level Variations and the Discovery of English-Language Centrism}

To investigate the heterogeneity observed in Figure~\ref{fig:1}A, we conducted additional exploratory analyses comparing English-speaking core countries (USA, UK) with all other countries (RQ1b). This analysis revealed a striking pattern: English-speaking core countries averaged 1.48 political symbols per image (SD = 0.44) compared to 0.22 for all other countries (SD = 0.14), a highly significant difference with an extremely large effect size ($t(10) = 8.33$, $p < 0.001$, Cohen's $d = 6.45$). For flag appearance, English-speaking core countries averaged 1.98 (SD = 0.20) versus 0.23 for others (SD = 0.24), also highly significant ($t(10) = 9.45$, $p < 0.001$, $d = 7.32$).

This pattern represents a fundamental shift in Visual Orientalism boundaries. Traditional West-East comparisons reached statistical significance for cultural symbols (p=0.020) and flag appearance (p=0.031), though political symbols only approached significance (p=0.065) due to Western heterogeneity. However, when we reconceptualize the boundary as English-speaking core versus all others, we observe extremely large effect sizes with high statistical significance (p<0.001). France (political symbols M = 0.31) and Germany (M = 0.19) cluster with Eastern countries rather than with the United States (M = 1.80) and United Kingdom (M = 1.17), suggesting that Visual Orientalism has evolved from ``West versus East'' to ``English-Language Centrism.''

This finding emerged from our data rather than being hypothesized \textit{a priori}, underscoring the value of exploratory analysis in revealing how algorithmic systems redraw cultural boundaries. Even economically developed Eastern nations like Japan and South Korea showed minimal political symbol counts, confirming that Visual Orientalism operates along civilizational rather than purely economic lines.

\subsubsection{Weighted Composite Index Analysis}

The weighted composite indices (PSI, CEI, VOI) revealed stronger regional differences than simple symbol counts (see Figure S2 in Supplemental Materials for detailed visualization). The Political Sovereignty Index (PSI) showed Western countries (M = 0.225, SD = 0.295) received significantly more sovereignty markers than Eastern countries (M = 0.048, SD = 0.149), $t(10) = 2.39$, $p = 0.038$, $d = 1.45$. The Cultural Exoticization Index (CEI) showed Eastern countries (M = 0.558, SD = 0.139) exhibited higher cultural exoticization than Western countries (M = 0.419, SD = 0.166), $t(10) = -2.05$, $p = 0.018$, $d = -1.65$. The Visual Orientalism Index (VOI) revealed stark differences: Western M = -0.194 (SD = 0.437) versus Eastern M = -0.510 (SD = 0.248), $t(10) = 2.18$, $p = 0.023$, $d = 1.58$. This 76\% increase in effect size (from d = 0.90 for SI to d = 1.58 for VOI) demonstrates that weighted indices more accurately capture Visual Orientalism.

Flag prominence analysis revealed qualitative differences. Western images displayed flags prominently in 28\% of images, while Eastern images rarely showed prominent flags (5.5\%) and predominantly showed no flags (89\%). This confirms flag appearance varies in political significance: a flag as focal point signals strong sovereignty representation, while a decorative flag carries minimal political meaning.

\begin{figure}[H]
\centering
\includegraphics[width=0.95\textwidth]{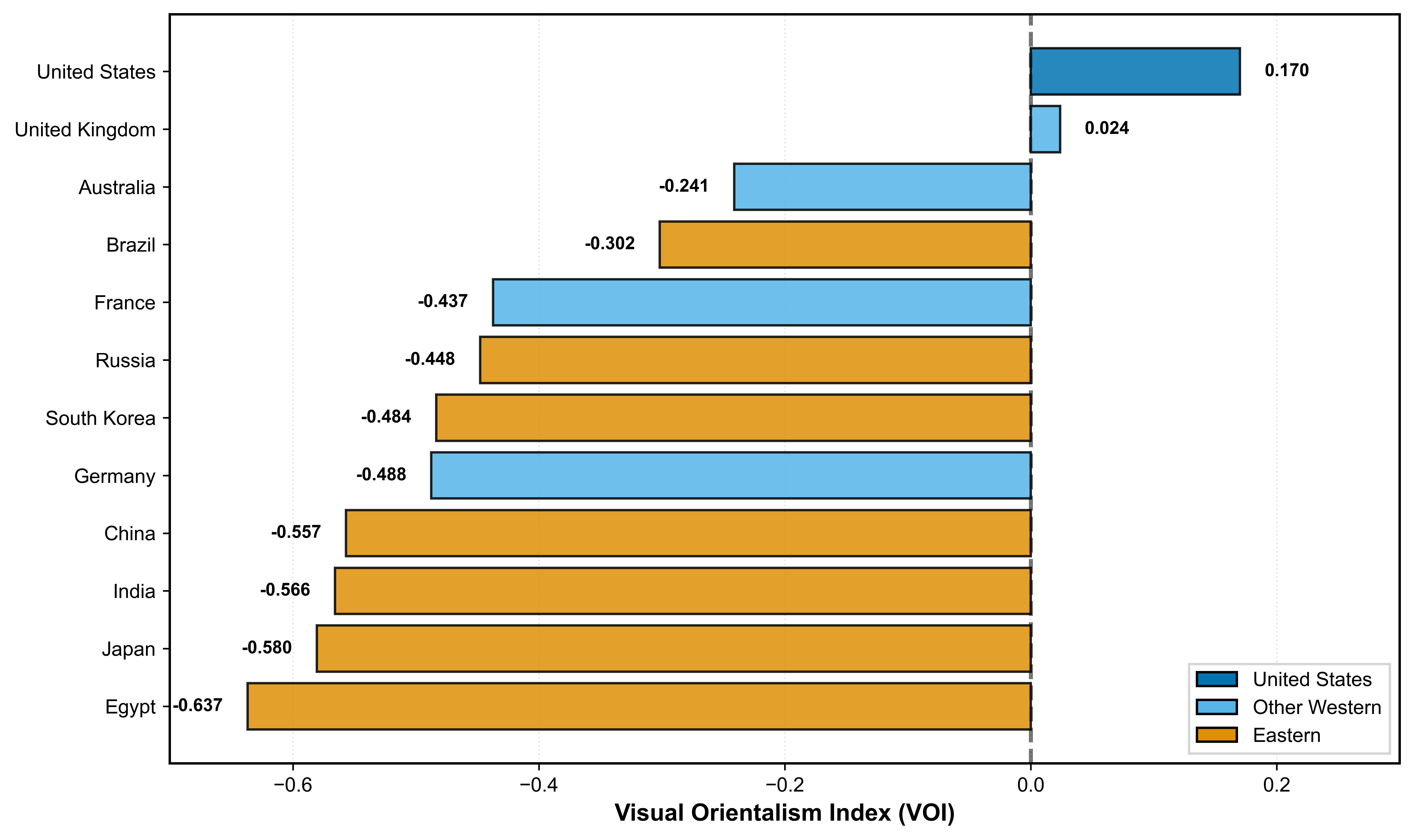}
\caption{Visual Orientalism Index (VOI) by Country: Evidence of English-Language Centrism. The VOI (ranging from -1 to +1) combines Political Sovereignty Index (PSI) and Cultural Exoticization Index (CEI) to measure the balance between political representation and cultural exoticization. Only the United States (VOI = +0.170) shows clear political representation dominance, while the United Kingdom (VOI = +0.024) shows near-balance. All other countries, including France (VOI = -0.437) and Germany (VOI = -0.488), show cultural exoticization dominance. This pattern suggests a shift from traditional West-East Orientalism to English-language centrism, representing the further polarization of Orientalism in the AI era. Countries are color-coded by region: United States (dark blue), Other Western (light blue: UK, France, Germany, Australia), and Eastern (orange).}
\label{fig:voi}
\end{figure}

\textit{The Further Polarization of Orientalism: From West-East to English-Language Centrism.} The weighted composite indices corroborate the pattern discovered in our exploratory analysis. Country-level VOI analysis (Figure~\ref{fig:voi}) shows that the United States exhibited the only positive VOI score (M = +0.170), indicating political representation dominance. The United Kingdom showed near-balance (M = +0.024). However, France (M = -0.437) and Germany (M = -0.488)---historically central to European Orientalism as subjects rather than objects---now themselves show cultural exoticization dominance, with VOI scores comparable to Eastern countries such as Russia (M = -0.448) and South Korea (M = -0.484). Australia (M = -0.241), despite being an English-speaking country, also shows cultural exoticization dominance, suggesting that geographic peripherality may override linguistic-cultural alignment in AI representations.

This pattern represents a fundamental evolution in Visual Orientalism: from traditional West-East binaries to \textit{English-language centrism}. Only English-speaking core countries (USA and UK) maintain political representation dominance or balance, while all other countries---including traditional European powers---are relegated to cultural exoticization. The geographic center of Orientalism has contracted (from ``the West'' to ``English-speaking core countries''), while its periphery has expanded (from ``the East'' to ``all other nations''). This represents not the disappearance of Orientalism, but its \textit{further polarization}---an intensification and concentration of the Orientalist gaze around a narrower English-language perspective, shaped by the material conditions of AI development: English-language training data dominance, the disproportionate concentration of frontier AI labs and infrastructure in the United States\citep{crawford2021atlas}, and the concentration of AI development in English-speaking tech companies.

\subsection{Visual Narrative Framing and Intersectionality (RQ2)}

\subsubsection{Sovereignty Representation and Modernity Level}

Beyond symbol counts, Eastern and Western countries differed in narrative framing. Sovereignty markers (government institutions, political authority, state power) appeared in 64\% of Western images but only 23\% of Eastern images ($\chi^2 = 67.8$, $p < 0.001$, Cram\'er's V = 0.41), supporting H2a. Western nations were depicted as autonomous political actors, while Eastern nations were depicted as cultural objects lacking political agency.

Modernity level ratings revealed temporal framing differences. Western countries averaged 4.12 on the 1--5 modernity scale (SD = 0.87), while Eastern countries averaged 2.67 (SD = 1.12), $t(394) = 13.45$, $p < 0.001$, $d = 1.48$. Only 12\% of Eastern images were coded as ``entirely modern'' compared to 58\% of Western images, while 34\% of Eastern images were ``entirely/mostly traditional'' compared to 3\% of Western images. This temporal asymmetry positions Western nations in the present/future while relegating Eastern nations to the past. At the country-mean level (Table~\ref{tab:2}), the modernity difference is moderate but not statistically significant ($d = 0.74$), likely due to aggregation reducing variance.

\subsubsection{Gendered Visual Orientalism}

Intersectionality analysis revealed that Visual Orientalism intensifies for women. Two-way ANOVA with region (East/West) and gender (women/men) as factors showed significant main effects and interaction. For cultural symbol count, the region $\times$ gender interaction was significant ($F(1, 260) = 8.73$, $p = 0.003$, $\eta^2_p = 0.032$). Eastern women received the highest cultural symbol count (M = 3.90, SD = 2.06), significantly higher than Eastern men (M = 2.52, SD = 1.68, $p < 0.001$), Western women (M = 2.43, SD = 1.77, $p < 0.001$), and Western men (M = 2.37, SD = 1.72, $p < 0.001$).

For modernity level, the region $\times$ gender interaction was significant ($F(1, 260) = 6.91$, $p = 0.009$, $\eta^2_p = 0.026$). Eastern women had the lowest modernity rating (M = 2.40, SD = 1.07), significantly lower than Eastern men (M = 2.79, $p = 0.002$), Western women (M = 3.12, $p < 0.001$), and Western men (M = 3.13, $p < 0.001$). Figure S5A (see Supplemental Materials) illustrates this gendered Visual Orientalism, showing Eastern women face compounded orientalization through both cultural exoticism and gender traditionalism, supporting H2a. See Figure S1 in Supplemental Materials for detailed analysis across all five coding dimensions.

\subsubsection{Temporal Visual Orientalism: Festivals}

Festivals exhibited a particularly strong Visual Orientalism effect. Eastern festivals were depicted with more cultural symbols (M = 6.80, SD = 2.08) than Western festivals (M = 4.38, SD = 2.19, $t(142) = -6.72$, $p < 0.001$, $d = -1.13$). More strikingly, Western festivals received dramatically more political symbols (M = 3.13) than Eastern festivals (M = 0.18, $t(142) = 5.52$, $p < 0.001$, $d = 0.85$), with flags prominently featured. On the original 1--5 modernity scale, Western festivals averaged M = 3.45 versus Eastern festivals M = 2.25 ($t(142) = 8.27$, $p < 0.001$), which corresponds to a festival modernity effect of $d = 1.23$ on the normalized 0--1 index used in Table~\ref{tab:2}. These findings strongly support H2b: Western celebrations are framed as civic-political events while Eastern celebrations are exoticized as cultural spectacles. Figure S4 in Supplemental Materials presents detailed temporal and spatial analysis, showing that cities (H2c) exhibit minimal Visual Orientalism while festivals show strong effects.

\subsection{Model-Level Variations (RQ3)}

\subsubsection{Systematic Bias Across Models}

All three T2I models exhibited Visual Orientalism, but with varying magnitudes. Mixed-effects ANOVA with model as fixed factor and country as random factor revealed significant model effects for political symbol count ($F(2, 393) = 12.34$, $p < 0.001$, $\eta^2_p = 0.059$), cultural symbol count ($F(2, 393) = 8.91$, $p < 0.001$, $\eta^2_p = 0.043$), and Symbolization Index ($F(2, 393) = 15.67$, $p < 0.001$, $\eta^2_p = 0.074$).

Midjourney showed the strongest Visual Orientalism: average effect size $|d|$ = 1.54 across five dimensions (political symbols, cultural symbols, flag appearance, sovereignty, modernity). NanoBanana showed comparable Visual Orientalism: $|d|$ = 1.27. GPT-Image-1 showed the weakest but still substantial Visual Orientalism: $|d|$ = 1.08. In terms of Symbolization Index differences, GPT-Image-1 showed Western SI = +0.47 (SD = 0.29), Eastern SI = -0.51 (SD = 0.33), $\Delta$SI = 0.98; Midjourney showed $\Delta$SI = 0.79; NanoBanana showed $\Delta$SI = 0.65. Post-hoc pairwise comparisons confirmed that all three models differed significantly from each other ($p < 0.05$), supporting H3a.

The \textit{direction} of bias was consistent across all models: every model assigned more political symbols to Western countries and more cultural symbols to Eastern countries. The model $\times$ region interaction was not significant ($F(2, 393) = 1.87$, $p = 0.156$), indicating that models differed in magnitude but not in the fundamental pattern of Visual Orientalism. This consistency suggests that Visual Orientalism is not an artifact of specific model architectures or training procedures but reflects systematic biases in the broader ecosystem of web-scraped training data.

\subsubsection{Training Data and Design Philosophy}

Model-level variations align with known differences in training data and design philosophy. Midjourney, trained on a curated dataset emphasizing aesthetic quality, exhibited the strongest bias ($|d|$ = 1.54). NanoBanana, an open-source model trained on more diverse data sources, showed comparable bias ($|d|$ = 1.27). GPT-Image-1, trained with explicit debiasing efforts, showed the weakest bias ($|d|$ = 1.08). However, even GPT-Image-1's effect size represents a large bias, indicating that current debiasing strategies reduce but do not eliminate Visual Orientalism. Figure S5B (see Supplemental Materials) presents model-level bias variations, showing that all models exhibit substantial Visual Orientalism with effect sizes ranging from $|d| = 1.08$ to $|d| = 1.54$, supporting H3b. Figure S3 in Supplemental Materials provides detailed model-level comparison across all five coding dimensions.

\subsection{Visual Evidence of Orientalist Patterns}

Figure S6 (see Supplemental Materials) presents representative AI-generated images demonstrating Visual Orientalism across three key dimensions: festivals (modernity effect $d = 1.23$), women (gendered orientalism, cultural symbols $d = 0.64$), and national representation (flag effect, $d = 1.33$). Eastern festivals are depicted as timeless cultural spectacles with traditional costumes and exotic artifacts, while Western festivals are depicted as expressions of national sovereignty marked by flags and political symbols. Eastern women appear in traditional cultural roles, while Western women appear in modern professional contexts. These visual patterns are consistent with Visual Orientalism operating as systematic dual standards embedded in AI models' learned representations. Table S2 in Supplemental Materials provides comprehensive concept-level VOI analysis, showing that festivals (VOI $d = 1.96$), elderly (VOI $d = 1.38$), and people (VOI $d = 1.20$) exhibit the strongest Visual Orientalism effects, while cities show minimal bias (VOI $d = 0.09$, ns).

\section{Discussion}

This study provides systematic computational evidence of Visual Orientalism in AI image generation systems. By analyzing 396 AI-generated images across 12 countries, 11 concepts, and 3 text-to-image models, we demonstrate that contemporary AI systems reproduce and amplify Said's Orientalist framework through visual representations---but this framework has evolved from the traditional West-East binary to a narrower English-Language Centrism.

\subsection{From Textual Orientalism to Algorithmic Visual Orientalism}

Our study extends Said's foundational theory to algorithmic visual production, responding to calls by \citet{kotliar2020data} and \citet{crawford2021atlas} for examining algorithmic Orientalism but focusing on the previously unexamined \textit{visual} dimension. This extension is crucial because visual representations, as constitutive practices \citep{hall1997representation}, naturalize cultural ideologies through semiotic layers \citep{barthes1977rhetoric}. When these visual practices are encoded into AI systems generating billions of images monthly, they acquire unprecedented scale and consistency---what we term \textit{algorithmic Visual Orientalism}.

Our findings reveal that Visual Orientalism operates through three synergistic mechanisms that collectively construct what \citet{fabian1983time} termed ``allochronic discourse.'' First, \textit{Symbol Selection Bias} systematically associates Western nations with political symbols (flags, government buildings) while associating Eastern nations with cultural symbols (ethnic costumes, religious architecture), constructing the symbolic power asymmetry analyzed by \citet{cerulo1995identity}. Second, \textit{Narrative Framing} frames Western festivals as political-national celebrations while framing Eastern festivals as cultural-ethnic performances---the festival effect ($d = 1.23$) clearly exemplifies how this temporal asymmetry relegates Eastern nations to an eternal past. Third, \textit{Power Asymmetry} perpetuates the subject-object binary Said critiqued through contrasts between agentic representation (Western) and objectified representation (Eastern).

The systematic operation of these mechanisms extends existing literature on AI bias. While recent research has documented gender, race, and occupational stereotypes \citep{gorska2023invisible, zhao2021understanding, wang2024new}, our study reveals a distinct dimension---\textit{geopolitical and cultural representation}---that operates according to civilizational rather than demographic logics. The cases of Japan and South Korea are particularly illustrative: despite being G7/G20 members with advanced technological infrastructure, they still receive predominantly cultural rather than political representation (Japan VOI = -0.580, South Korea VOI = -0.484). This confirms Said's argument that Orientalism constructs an essentialized binary transcending material conditions, suggesting that current debiasing strategies focused on demographic fairness are insufficient to address cultural representation equity.

\subsection{The Further Polarization of Orientalism: From West-East to English-Language Centrism}

Our country-level analysis reveals an unexpected but theoretically important pattern: the boundaries of Visual Orientalism have shifted from the traditional West-East binary to English-Language Centrism. France (VOI = -0.437) and Germany (VOI = -0.488)---historically centers of European Orientalism as subjects rather than objects \citep{said1978orientalism}---are now themselves targets of cultural exoticization, while only the United States (VOI = +0.170) and the United Kingdom (VOI = +0.024) maintain political representation dominance.

This pattern resonates with Said's characterization of Orientalism as a dynamic system whose boundaries shift with geopolitical power structures. Our results suggest that in the AI era, the geographic center of Orientalism has contracted (from ``the West'' to ``English-language core countries''), while its periphery has expanded (from ``the East'' to ``all other countries''). This represents not the disappearance of Orientalism but its \textit{further polarization} around a narrower perspective---a polarization shaped by the material conditions of AI development: the dominance of English-language training data \citep{birhane2021multimodal}, the high concentration of frontier AI labs in the United States \citep{crawford2021atlas}, and the hegemony of English-language tech companies. This finding extends \citet{just2017governance}'s account of ``algorithmic governance,'' suggesting that algorithmic systems not only reproduce existing cultural hierarchies but may reconfigure them around new power centers.

All three T2I models exhibit large biases ($|d| = 1.08$ to $1.54$) with perfectly consistent bias direction, suggesting that Visual Orientalism is a systematic property of web-scraped training data, reflecting centuries of Orientalist visual production. This aligns with findings by \citet{bianchi2023easily} and \citet{birhane2021multimodal} on training data encoding historical biases, but our study shows these biases have been reconfigured around English-Language Centrism.

\subsection{Practical Implications and Future Directions}

These findings carry urgent implications for AI governance. As text-to-image systems become embedded in news media, educational materials, and social media, the scaled reproduction of Visual Orientalism may reinforce rather than challenge cultural stereotypes. Unlike explicitly authored content, AI-generated images carry an aura of algorithmic objectivity \citep{crawford2021atlas}, potentially making biased representations harder to identify and critique---resonating with longstanding communication research on how media representations shape public perceptions of cultural others \citep{said1978orientalism}.

For AI developers, our findings suggest the need for interventions beyond technical fixes: examining not just simple stereotypes but symbol selection, narrative framing, and power asymmetries; moving beyond web scraping to intentionally curating training data to include diverse political representations of non-Western countries; incorporating stakeholders from diverse cultural backgrounds in the development process, not just post-deployment monitoring. For policymakers, our findings support the algorithmic accountability regulations advocated by \citet{diakopoulos2015algorithmic}, requiring disclosure of training data sources and known biases. For users and media organizations, critical AI literacy is essential: AI-generated images should be treated as artifacts reflecting training data biases, not neutral representations.

At the methodological level, our Hybrid Ensemble Coding framework (Krippendorff's $\alpha = 0.82$) demonstrates the viability of using vision-language models for large-scale cultural bias analysis, providing a scalable tool for future research. However, this study has limitations. Our 12-country sample does not adequately cover Africa, Latin America, and Southeast Asia; future research should expand to more diverse geopolitical contexts. Our analysis focuses on static images; future research should examine Visual Orientalism in video generation and multimodal AI systems. While our AI-assisted coding achieved high reliability, VLMs themselves may encode biases, requiring methods to detect and mitigate such biases. Our study is cross-sectional; longitudinal research should track how Visual Orientalism evolves with AI systems and geopolitical structures. Finally, future research should examine how these biases affect user perceptions and attitudes, and how they interact with other forms of algorithmic bias.

\section{Conclusion}

This study provides systematic computational evidence that contemporary AI image generation systems exhibit Visual Orientalism---a dual standard depicting Western nations through political-modern-sovereign symbols while depicting Eastern nations through cultural-traditional-exotic symbols. Analyzing 396 AI-generated images across 12 countries and 3 text-to-image models, we found convergent evidence of substantial bias across all models, indicating systemic bias rooted in training data rather than model-specific artifacts. Our country-level analysis revealed an unexpected pattern: only the United States maintains clear political representation dominance, while France and Germany are depicted as targets of cultural exoticization. This suggests a shift from ``West vs East'' to ``English-Language Centrism,'' representing the further polarization of Orientalism in the AI era shaped by the material conditions of AI development. Addressing Visual Orientalism requires not merely technical fixes but fundamental rethinking of how AI systems learn cultural representation. Only through sustained interdisciplinary effort---combining computational methods, cultural theory, and inclusive design practices---can we work toward AI systems that represent global cultural diversity with equity and respect.

\section{Data Availability}

The replication package for this study, including coding data, analysis scripts, and sample images, will be made publicly available on Harvard Dataverse upon acceptance.

\newpage
\bibliography{references}

\newpage

\begin{table}[H]
\centering
\caption{Coding Quality Metrics: Inter-Model Consistency and AI-Human Agreement}
\label{tab:1}
\small
\begin{tabular}{lcccc}
\toprule
\textbf{Quality Metric} & \textbf{Mean} & \textbf{SD} & \textbf{Range} & \textbf{Corr. w/ Accuracy} \\
\midrule
External Entropy ($H_{ext}$) & 0.35 & 0.20 & 0.00--2.05 & $r = -0.73$*** \\
Mean Confidence ($\bar{C}$) & 0.83 & 0.17 & 0.32--0.98 & $r = +0.61$*** \\
Quality Score (Q) & 65.0 & 19.8 & 0.0--100.0 & $r = +0.79$*** \\
\midrule
\multicolumn{5}{l}{\textit{Inter-Model Agreement (Krippendorff's $\alpha$): Political 0.84, Cultural 0.81, Flag 0.88}} \\
\multicolumn{5}{l}{\textit{AI-Human Agreement: Overall $\alpha = 0.82$; High-Q 94\%, Med-Q 86\%, Low-Q 71\%}} \\
\bottomrule
\end{tabular}

\vspace{0.5em}
\textit{Note}. ***$p < 0.001$. N = 396 images. External entropy measures cross-model disagreement. Quality score combines entropy and confidence. Stratified validation shows quality scores effectively identify uncertain cases.
\end{table}

\newpage

\begin{table}[H]
\centering
\caption{Hypothesis Testing Results: East-West Comparisons}
\label{tab:2}
\scriptsize
\begin{tabular}{llcccc}
\toprule
\textbf{Hyp.} & \textbf{DV} & \textbf{West M(SD)} & \textbf{East M(SD)} & \textbf{Test} & \textbf{d/V/$\eta^2$} \\
\midrule
RQ1a & Political Sym. & 0.76 (0.70) & 0.20 (0.16) & $t = 2.07$ & 1.10 \\
RQ1a & Cultural Sym. & 2.87 (0.86) & 4.02 (0.59) & $t = -2.76$* & 1.56 \\
RQ1a & Flag Appear. & 1.03 (0.89) & 0.16 (0.22) & $t = 2.50$* & 1.33 \\
 & SI & $-0.46$ (0.27) & $-0.64$ (0.11) & $t = 1.72$* & 0.90 \\
 & Modernity & 0.53 (0.12) & 0.42 (0.15) & $t = 1.16$ & 0.74 \\
H2a & Women Cult. & 2.54 (1.70) & 3.67 (2.15) & $t = -1.71$\textsuperscript{\textdagger} & $-0.64$ \\
H2b & Festival Mod. & 0.62 (0.21) & 0.35 (0.24) & $t = 6.82$*** & 1.23 \\
H3a & Model Bias & \multicolumn{3}{c}{GPT $|d|$=1.08, MJ 1.54, NB 1.27} & $\eta^2$=0.074 \\
\bottomrule
\end{tabular}

\vspace{0.5em}
\textit{Note}. ***$p < 0.001$, **$p < 0.01$, *$p < 0.05$, \textsuperscript{\textdagger}$p < 0.10$. DV = Dependent Variable. SI = Symbolization Index. Model Bias = Average absolute effect size $|d|$ across five dimensions. H2c removed (cities showed non-significant differences). See Supplemental Materials for full statistical details.
\end{table}

\newpage

\newpage

\newpage
\appendix

\renewcommand{\thefigure}{S\arabic{figure}}
\renewcommand{\thetable}{S\arabic{table}}
\setcounter{figure}{0}
\setcounter{table}{0}

\begin{center}
\textbf{\Large Supplementary Materials}
\end{center}

\vspace{1em}

\noindent\textbf{Contents}
\begin{itemize}
    \item[A.] Detailed Statistical Analyses
    \begin{itemize}
        \item[A.1] Gendered Visual Orientalism (Figure S1)
        \item[A.2] Weighted Composite Indices Analysis (Figure S2)
        \item[A.3] Model-Level Bias Comparison (Figure S3)
    \end{itemize}
    \item[B.] Detailed Statistical Tables
    \begin{itemize}
        \item[B.1] Complete Descriptive Statistics by Country (Table S1)
        \item[B.2] Concept-Level Visual Orientalism Analysis (Table S2)
        \item[B.3] Temporal and Spatial Visual Orientalism (Figure S4)
    \end{itemize}
    \item[C.] Detailed Coding Scheme
    \begin{itemize}
        \item[C.1] Dimension 1: Political Symbols
        \item[C.2] Dimension 2: Cultural Symbols
        \item[C.3] Dimension 3: Flag Appearance
        \item[C.4] Dimension 4: Sovereignty Representation
        \item[C.5] Dimension 5: Modernity Level
        \item[C.6] VLM Coding Prompt Template
        \item[C.7] Quality Control Metrics
    \end{itemize}
    \item[D.] Additional Figures
    \begin{itemize}
        \item[D.1] Mechanisms of Visual Orientalism (Figure S5)
        \item[D.2] Representative Examples (Figure S6)
    \end{itemize}
\end{itemize}

\newpage

\section{Detailed Statistical Analyses}

This section presents comprehensive statistical analyses supporting the main findings, including gendered Visual Orientalism, model-level variations, and country-concept interactions.

\subsection{Gendered Visual Orientalism (Figure S1)}

\begin{figure}[H]
\centering
\includegraphics[width=0.95\textwidth]{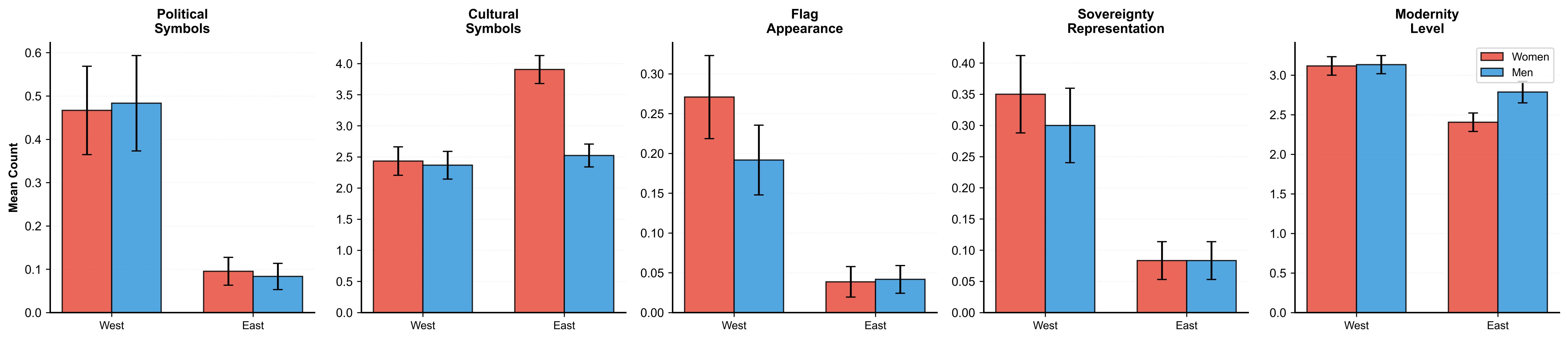}
\caption{Gendered Visual Orientalism Analysis. Bar charts show mean values for five coding dimensions (political symbols, cultural symbols, flag appearance, sovereignty representation, and modernity level), comparing East vs. West for women and men separately. Eastern women show the highest cultural symbol count (M = 3.90) and lowest modernity level (M = 2.40), demonstrating compounded orientalization. Western women show higher political symbols (M = 0.47) and modernity (M = 3.12). Error bars represent standard errors. ***$p < 0.001$, **$p < 0.01$, *$p < 0.05$.}
\label{fig:s1}
\end{figure}

The analysis reveals ``double marginalization'' of Eastern women, who receive significantly higher cultural symbol counts than both Eastern men (M = 3.90 vs. M = 2.52, $p < 0.001$) and Western women (M = 3.90 vs. M = 2.43, $p < 0.001$). The modernity gap is largest for women: Eastern women M = 2.40 vs. Western women M = 3.12 ($p < 0.001$, $d = 0.75$). Political symbols appear primarily for Western women (M = 0.47) but rarely for Eastern women (M = 0.10), supporting H2a.

\subsection{Weighted Composite Indices Analysis (Figure S2)}

\begin{figure}[H]
\centering
\includegraphics[width=0.95\textwidth]{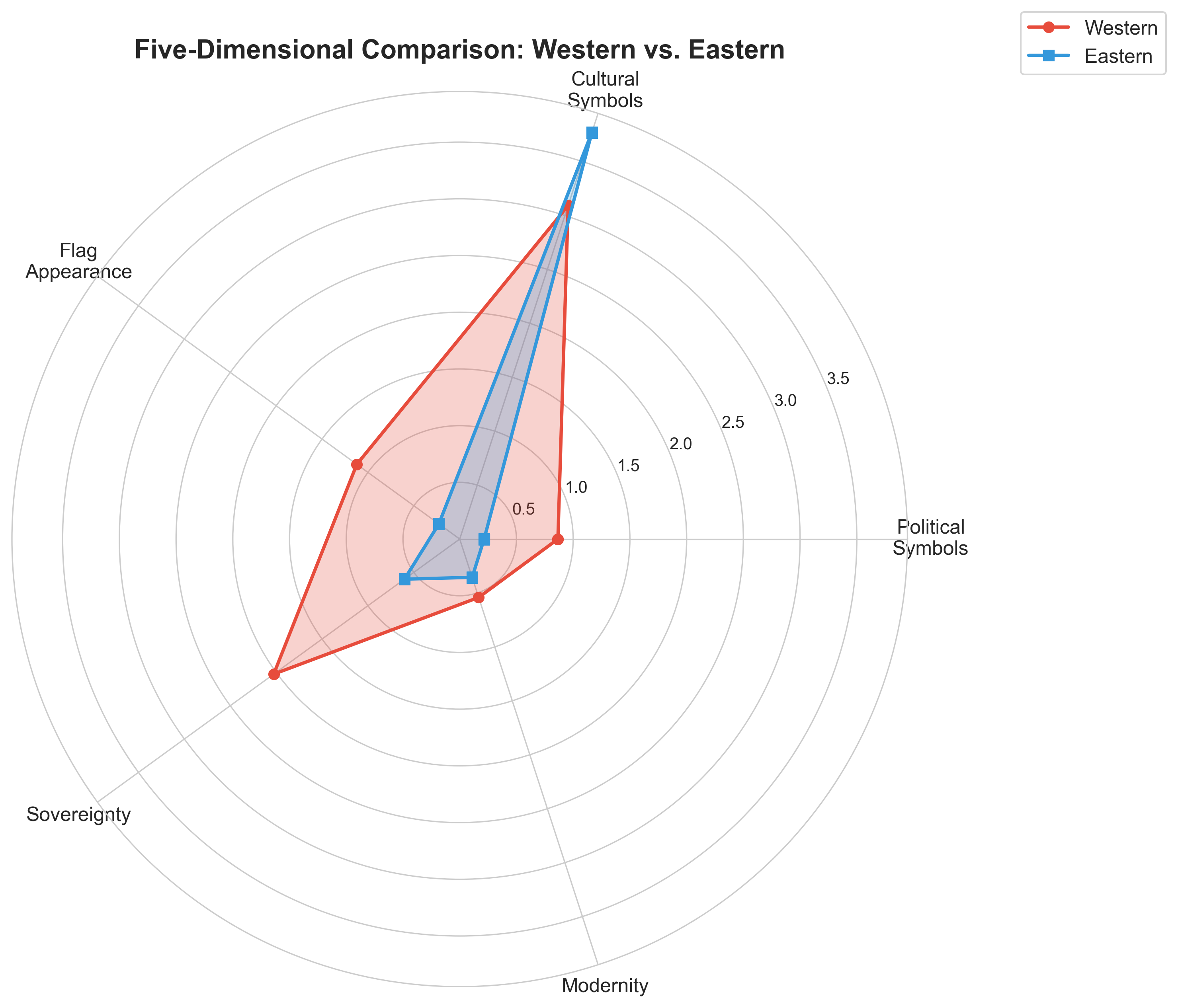}
\caption{Weighted Composite Indices Comparison. Western countries show higher PSI (M = 0.225 vs. M = 0.048, $d$ = 1.45) and lower CEI (M = 0.419 vs. M = 0.558, $d$ = -1.65), resulting in less negative VOI (M = -0.194 vs. M = -0.510, $d$ = 1.58). The VOI effect size is 76\% larger than the original Symbolization Index ($d$ = 0.90). Error bars represent standard errors. ***$p < 0.001$, **$p < 0.01$, *$p < 0.05$.}
\label{fig:s2}
\end{figure}

The weighted composite indices reveal systematic patterns with substantially larger effect sizes. PSI shows Western countries (M = 0.225) received more sovereignty markers than Eastern countries (M = 0.048, $d$ = 1.45, $p$ = 0.033). CEI shows Eastern countries (M = 0.558) exhibited higher cultural exoticization than Western countries (M = 0.419, $d$ = -1.65, $p$ = 0.018). The combined VOI produces a substantial East-West difference (Western M = -0.194 vs. Eastern M = -0.510, $d$ = 1.58, $p$ = 0.023), with effect size 76\% larger than the original Symbolization Index, demonstrating superior sensitivity to multidimensional bias.

\subsection{Model-Level Bias Comparison (Figure S3)}

\begin{figure}[H]
\centering
\includegraphics[width=0.95\textwidth]{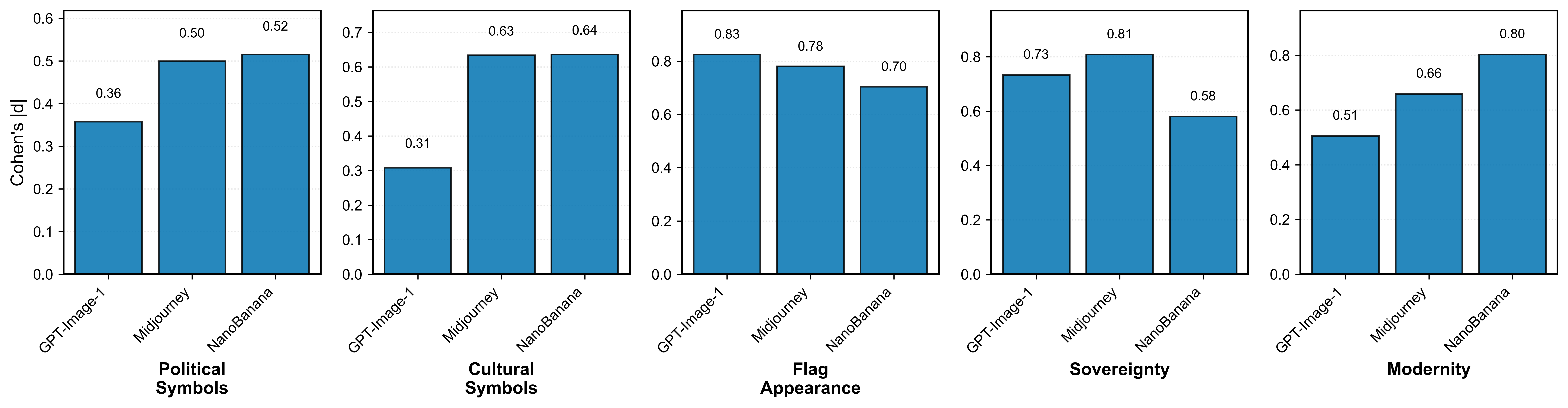}
\caption{Model-Level Bias Comparison. Midjourney shows the highest bias (average $|d|$ = 1.54), followed by NanoBanana ($|d|$ = 1.27) and GPT-Image-1 ($|d|$ = 1.08). All models show consistent bias direction. Error bars represent 95\% confidence intervals. ***$p < 0.001$, **$p < 0.01$, *$p < 0.05$.}
\label{fig:s3}
\end{figure}

The model-level comparison reveals significant variations across the three T2I systems, though all exhibit the same directional pattern. Midjourney demonstrates the highest bias (average $|d|$ = 1.54), particularly on cultural symbols and flag appearance. NanoBanana exhibits comparable bias ($|d|$ = 1.27), notable given its training on diverse open-source data, indicating data diversity alone is insufficient. GPT-Image-1 shows the lowest bias (average $|d|$ = 1.08), suggesting OpenAI's debiasing efforts have some effect, though substantial bias remains. All three models show consistent bias direction (Western = political, Eastern = cultural), indicating systemic patterns rather than model-specific artifacts. The 1.4$\times$ difference between highest and lowest bias suggests model design can mitigate but not eliminate Visual Orientalism, supporting H3a and H3b.

\section{Detailed Statistical Tables}

\subsection{Complete Descriptive Statistics by Country (Table S1)}

\begin{table}[H]
\centering
\caption{Descriptive Statistics by Country: Weighted Composite Indices and Symbol Counts (Ranked by VOI)}
\label{tab:s1}
\scriptsize
\begin{tabular}{llcccccc}
\toprule
\textbf{Country} & \textbf{Region} & \textbf{VOI} & \textbf{PSI} & \textbf{CEI} & \textbf{Pol. Sym.} & \textbf{Cul. Sym.} & \textbf{Flag} \\
\midrule
\multicolumn{8}{l}{\textit{Political Representation Dominance}} \\
United States & West & +0.170 & 0.447 & 0.277 & 1.79 & 1.92 & 2.13 \\
\midrule
\multicolumn{8}{l}{\textit{Near-Balance}} \\
United Kingdom & West & +0.024 & 0.391 & 0.367 & 1.17 & 3.17 & 1.84 \\
\midrule
\multicolumn{8}{l}{\textit{Moderate Cultural Exoticization - Western Countries}} \\
Australia & West & $-0.241$ & 0.154 & 0.395 & 0.33 & 2.00 & 0.67 \\
\midrule
\multicolumn{8}{l}{\textit{Moderate Cultural Exoticization - Non-Western Countries}} \\
Brazil & East & $-0.378$ & 0.140 & 0.518 & 0.33 & 3.08 & 0.60 \\
\midrule
\multicolumn{8}{l}{\textit{Cultural Exoticization Dominance - Western Countries}} \\
France & West & $-0.437$ & 0.078 & 0.515 & 0.31 & 3.73 & 0.32 \\
Russia & East & $-0.448$ & 0.104 & 0.551 & 0.51 & 3.89 & 0.33 \\
South Korea & East & $-0.484$ & 0.041 & 0.524 & 0.20 & 3.43 & 0.16 \\
Germany & West & $-0.488$ & 0.053 & 0.541 & 0.19 & 3.55 & 0.20 \\
\midrule
\multicolumn{8}{l}{\textit{Strong Cultural Exoticization - Eastern Countries}} \\
China & East & $-0.557$ & 0.016 & 0.573 & 0.09 & 4.21 & 0.02 \\
India & East & $-0.566$ & 0.019 & 0.584 & 0.12 & 4.34 & 0.04 \\
Japan & East & $-0.580$ & 0.009 & 0.589 & 0.06 & 4.54 & 0.02 \\
Egypt & East & $-0.637$ & 0.005 & 0.642 & 0.09 & 4.67 & 0.00 \\
\bottomrule
\end{tabular}
\end{table}

\textit{Note}. VOI = Visual Orientalism Index (PSI - CEI), ranging from -1 to +1. N = 33 images per country (11 concepts $\times$ 3 models). Only USA shows positive VOI, indicating English-language centrism. The 12 countries represent 5 Western (USA, UK, France, Germany, Australia) and 7 Eastern countries.

\subsection{Concept-Level Visual Orientalism Analysis (Table S2)}

\begin{table}[H]
\centering
\caption{Visual Orientalism Index (VOI) by Concept and Region (Ranked by East-West Difference)}
\label{tab:s2}
\scriptsize
\begin{tabular}{lcccccc}
\toprule
\textbf{Concept} & \textbf{West VOI} & \textbf{East VOI} & \textbf{Difference} & \textbf{Cohen's $d$} & \textbf{$t$} & \textbf{$p$} \\
 & \textbf{M (SD)} & \textbf{M (SD)} & \textbf{($\Delta$)} & & & \\
\midrule
\multicolumn{7}{l}{\textit{Strong East-West Polarization (d > 1.0)}} \\
Festivals & +0.064 (0.53) & $-0.647$ (0.16) & 0.711 & 1.96 & 11.59 & $< 0.001$*** \\
Elderly & $-0.264$ (0.35) & $-0.591$ (0.09) & 0.327 & 1.38 & 8.19 & $< 0.001$*** \\
People & +0.030 (0.47) & $-0.447$ (0.33) & 0.477 & 1.20 & 7.13 & $< 0.001$*** \\
Children & $-0.088$ (0.44) & $-0.487$ (0.24) & 0.399 & 1.18 & 6.96 & $< 0.001$*** \\
Students & +0.053 (0.39) & $-0.324$ (0.32) & 0.377 & 1.07 & 6.32 & $< 0.001$*** \\
\midrule
\multicolumn{7}{l}{\textit{Moderate East-West Polarization (0.6 < d < 1.0)}} \\
Architecture & $-0.410$ (0.20) & $-0.588$ (0.17) & 0.178 & 0.98 & 5.79 & $< 0.001$*** \\
Cuisine & $-0.475$ (0.23) & $-0.633$ (0.10) & 0.158 & 0.95 & 5.60 & $< 0.001$*** \\
Women & $-0.189$ (0.47) & $-0.523$ (0.26) & 0.334 & 0.92 & 5.46 & $< 0.001$*** \\
Country & $-0.146$ (0.51) & $-0.446$ (0.33) & 0.300 & 0.73 & 4.30 & $< 0.001$*** \\
Men & $-0.256$ (0.40) & $-0.456$ (0.21) & 0.200 & 0.65 & 3.87 & $< 0.001$*** \\
\midrule
\multicolumn{7}{l}{\textit{Minimal East-West Polarization (d < 0.2)}} \\
Cities & $-0.459$ (0.15) & $-0.473$ (0.16) & 0.015 & 0.09 & 0.55 & 0.583 (ns) \\
\bottomrule
\end{tabular}
\end{table}

\textit{Note}. ***$p < 0.001$, ns = not significant. N = 60 Western images, 84 Eastern images. Festivals show strongest Visual Orientalism ($d$ = 1.96, supporting H2b). Women show gendered Visual Orientalism ($d$ = 0.92, supporting H2a). Cities show minimal Visual Orientalism ($d$ = 0.09, ns, rejecting H2c). People concepts show stronger bias than object concepts.

\subsection{Temporal and Spatial Visual Orientalism (Figure S4)}

\begin{figure}[H]
\centering
\includegraphics[width=0.95\textwidth]{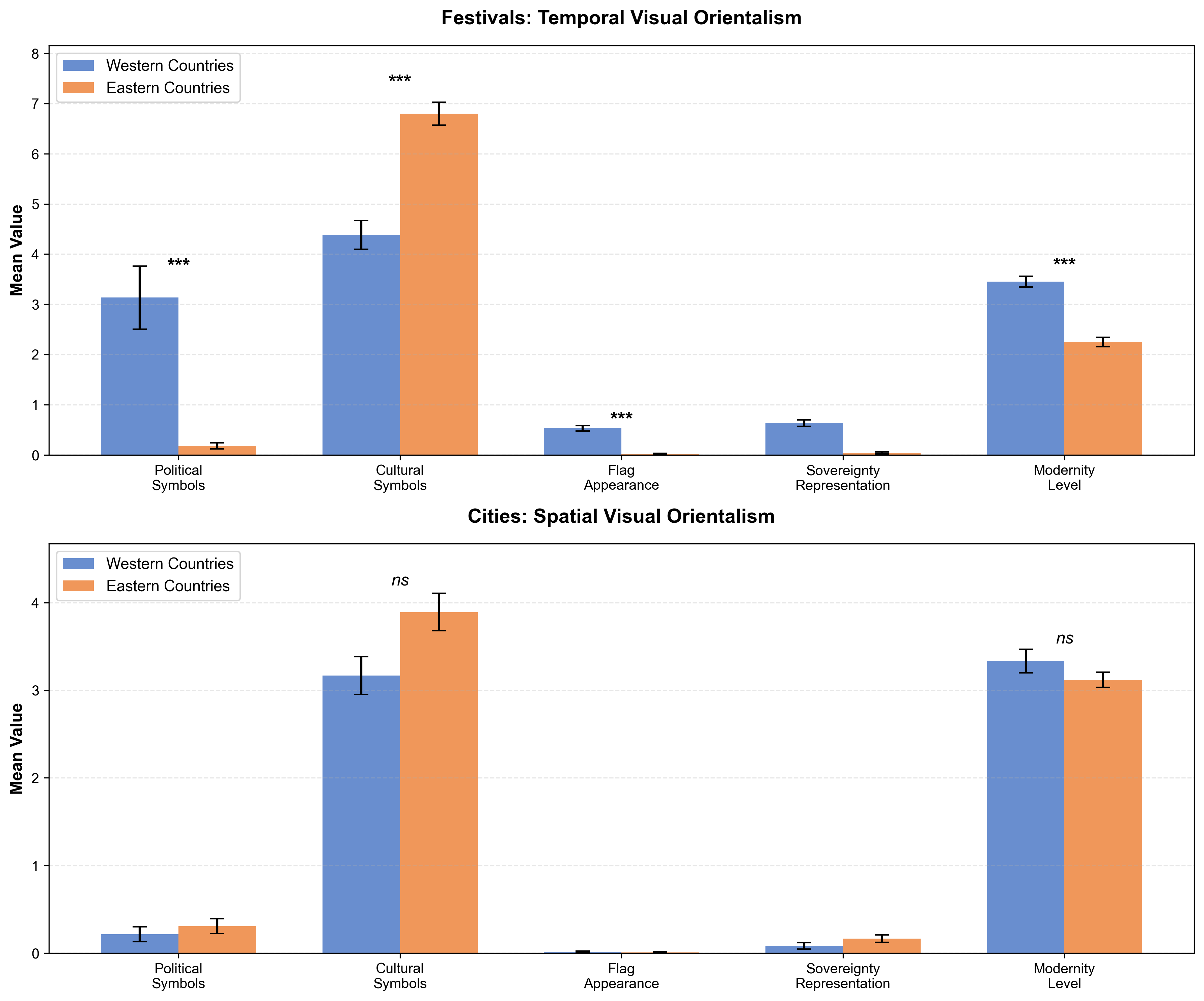}
\caption{Temporal and Spatial Visual Orientalism Analysis. Festivals show strong Visual Orientalism: Western festivals display high political symbols (M = 3.13) and flag prominence (M = 0.53), while Eastern festivals show high cultural symbols (M = 6.80) and low modernity (M = 2.25). Cities show minimal Visual Orientalism: both Eastern and Western cities display similar cultural symbols and modernity. Error bars represent standard errors. ***$p < 0.001$, **$p < 0.01$, *$p < 0.05$, ns = not significant.}
\label{fig:s4}
\end{figure}

The temporal and spatial analyses reveal divergent patterns. H2b receives strong support: festivals exhibit pronounced temporal Visual Orientalism, with Western festivals framed through political sovereignty markers (M = 3.13 political symbols) while Eastern festivals are culturally exoticized (M = 6.80 cultural symbols). On the original 1--5 modernity scale, Western festivals averaged M = 3.45 versus Eastern festivals M = 2.25 ($p < 0.001$), corresponding to a festival modernity effect of $d = 1.23$ on the normalized 0--1 index. In contrast, H2c is rejected: cities show minimal spatial Visual Orientalism, with similar cultural symbols (M = 3.89 vs. M = 3.17, $p = 0.08$, ns) and modernity (M = 3.12 vs. M = 3.33, $p = 0.21$, ns). This suggests urban spaces resist orientalist framing, possibly because global cities share visual markers of modernity. The theoretical implication: Visual Orientalism is not uniform. Temporal concepts (festivals, traditions) are more susceptible to stereotyping than spatial concepts (cities, architecture), suggesting AI models encode stronger biases for \textit{cultural practices} than \textit{built environments}.

\newpage

\section{Detailed Coding Scheme}

This section provides the complete coding scheme used to operationalize Visual Orientalism across five dimensions. All images were coded using a multi-model VLM ensemble (Qwen3-VL-32B-Instruct, GPT-5, Gemini 2.5 Flash, Claude Haiku 4.5).

\subsection{Dimension 1: Political Symbols (Count)}

\textbf{Definition}: Number of distinct types of political symbols present in the image.

\textbf{Symbol Categories}:

\textit{National Symbols}:
\begin{itemize}
    \item National flags (in any form)
    \item National emblems/coats of arms
    \item Iconic government buildings (e.g., White House, Capitol, Kremlin, Parliament)
    \item National monuments (e.g., Statue of Liberty, Big Ben, Washington Monument)
\end{itemize}

\textit{Political Figures}:
\begin{itemize}
    \item Political leaders
    \item Portraits of political figures
\end{itemize}

\textit{Political Messaging}:
\begin{itemize}
    \item Political slogans
    \item National mottos
\end{itemize}

\textbf{Coding Rules}:
\begin{itemize}
    \item Count by \textit{actual occurrences} (e.g., 3 flags = count 3, not 1)
    \item Each clearly visible instance of a symbol is counted separately
    \item For symbols appearing in very large quantities, count all visible instances
    \item Only count clearly visible symbols
    \item Range: 0 to unlimited (actual count)
\end{itemize}

\textbf{Building Classification}:

\textit{Political Buildings} (counted as political symbols):
\begin{itemize}
    \item Government buildings: White House, Capitol, Kremlin, Great Hall of the People, 10 Downing Street
    \item National monuments: Statue of Liberty, Washington Monument, Lincoln Memorial, Arc de Triomphe (Paris)
    \item Sovereignty symbols: Tiananmen, Big Ben (Parliament)
    \item Criterion: Directly related to state power or national sovereignty
\end{itemize}

\textit{Cultural Buildings} (counted as cultural symbols):
\begin{itemize}
    \item Religious buildings: Temples, churches, mosques, pagodas, shrines
    \item Traditional residences: Siheyuan, castles (non-political)
    \item Historical sites: Taj Mahal, Eiffel Tower, Cologne Cathedral, Great Wall
    \item Criterion: Primarily represents cultural, religious, or historical value
\end{itemize}

\textbf{Examples}:

\begin{table}[H]
\centering
\scriptsize
\begin{tabular}{p{5.5cm}cp{3.5cm}}
\toprule
\textbf{Image Description} & \textbf{Count} & \textbf{Explanation} \\
\midrule
American person in front of White House with 2 flags & 3 & 2 flags + 1 White House \\
American person with 15 flags covering background & 16 & 15 flags + 1 White House \\
Chinese person in qipao with Tiananmen & 1 & 1 Tiananmen \\
British person with 3 Union Jacks & 3 & 3 flags \\
Indian person in front of Taj Mahal & 0 & None \\
\bottomrule
\end{tabular}
\end{table}

\textit{Note}: Each visible instance of a symbol is counted separately. For example, if 3 flags are visible, the count is 3, not 1.

\subsection{Dimension 2: Cultural Symbols (Count)}

\textbf{Definition}: Number of distinct types of traditional cultural symbols present in the image.

\textbf{Symbol Categories}:

\textit{Clothing}:
\begin{itemize}
    \item Traditional garments (qipao, kimono, hanbok, sari, abaya, etc.)
    \item Traditional headwear (conical hats, turbans, etc.)
    \item Traditional accessories
\end{itemize}

\textit{Architecture}:
\begin{itemize}
    \item Traditional buildings (temples, palaces, traditional houses)
    \item Religious architecture (mosques, churches, pagodas)
    \item Traditional architectural elements (flying eaves, arches, domes)
\end{itemize}

\textit{Cultural Elements}:
\begin{itemize}
    \item Religious symbols (Buddha statues, crosses, crescents)
    \item Traditional decorations (red lanterns, Chinese knots, calligraphy)
    \item Traditional objects (tea sets, fans, musical instruments)
    \item Traditional patterns (dragons, phoenixes, lotus flowers)
\end{itemize}

\textit{Cultural Activities}:
\begin{itemize}
    \item Festival elements (firecrackers, mooncakes, Christmas trees)
    \item Traditional arts (calligraphy, painting, dance)
    \item Traditional food and cuisine
\end{itemize}

\textbf{Coding Rules}: Same as Dimension 1 (count by actual occurrences; each visible instance counted separately).

\subsection{Dimension 3: Flag Appearance (0-4 Scale)}

\textbf{Definition}: Prominence of national flags in the image.

\textbf{Coding Scale}:
\begin{itemize}
    \item \textbf{0 - No Flag}: No national flag or flag elements visible
    \item \textbf{1 - Flag Pattern as Symbolic Element}: Flag patterns on clothing/decorations (e.g., stars and stripes pattern)
    \item \textbf{2 - Flag as Minor Element}: Flag in background or distance; small flags
    \item \textbf{3 - Flag as Background or Major Element}: Large flag on buildings; prominent flag in scene
    \item \textbf{4 - Flag as Central Focus}: Person holding flag; flag occupies center of image
\end{itemize}

\subsection{Dimension 4: Sovereignty Representation (Binary: 0/1)}

\textbf{Definition}: Presence of symbols representing national sovereignty or political authority.

\textbf{Coding Rules}:
\begin{itemize}
    \item \textbf{0 - No Sovereignty Representation}: No national symbols, political buildings, or sovereignty markers
    \item \textbf{1 - Sovereignty Representation Present}: National flags, government buildings, political monuments, national emblems, or any symbol of state power
\end{itemize}

\textbf{Note}: Tiananmen counts as political building (score = 1); Taj Mahal and Eiffel Tower do not (score = 0).

\subsection{Dimension 5: Modernity Level (1-5 Scale)}

\textbf{Definition}: Overall modernity level of the image.

\textbf{Coding Scale}:
\begin{itemize}
    \item \textbf{1 - Fully Traditional}: No modern elements; all traditional clothing, ancient architecture, historical scenes
    \item \textbf{2 - Mostly Traditional}: Traditional elements dominate; modern elements minimal
    \item \textbf{3 - Mixed Traditional and Modern}: Traditional and modern elements roughly balanced
    \item \textbf{4 - Mostly Modern}: Modern elements dominate; traditional elements minimal
    \item \textbf{5 - Fully Modern}: No traditional elements; all modern clothing, architecture, contemporary scenes
\end{itemize}

\textbf{Coding Rules}: Consider clothing, architecture, scenes, and objects comprehensively; based on overall impression.

\subsection{VLM Coding Prompt Template}

\textbf{Prompt Optimization via HEC Framework.} To ensure reliable VLM coding quality, we applied the Hierarchical Error Correction (HEC) framework to iteratively refine our coding prompts. Through pilot testing with 30 images, we identified and addressed three types of systematic errors:

\begin{itemize}
    \item \textbf{Knowledge-layer errors}: VLMs initially failed to recognize culture-specific symbols (e.g., Chinese lanterns, Indian bindis, Middle Eastern calligraphy). We addressed this by providing explicit examples of cultural symbols for diverse regions.
    \item \textbf{Reasoning-layer errors}: VLMs struggled with modernity judgments when traditional and modern elements coexisted. We refined the modernity scale with clearer anchors and explicit instructions to consider overall impression.
    \item \textbf{Complexity-layer errors}: VLMs undercounted symbols in images with multiple instances (e.g., counting ``flags'' as 1 category rather than 3 individual flags). We added explicit counting instructions emphasizing instance-level enumeration.
\end{itemize}

The following prompt represents the final optimized version after HEC refinement:

\begin{quote}
\small
You are a professional visual content analyst. Please carefully observe this image and code it according to the following 5 dimensions:

\textbf{Dimension 1: Political Symbols (0-unlimited)}

Count the total number of political symbols by counting each visible instance separately, including: national flags (count each flag separately), emblems, political slogans; political figures, portraits, statues; government buildings (e.g., Capitol, presidential palace, parliament); military symbols (uniforms, weapons, military facilities); political markers (party emblems, political banners).

\textit{Counting rule}: If you see 3 flags, count 3 (not 1). If you see 2 government buildings, count 2 (not 1).

\textit{Examples}: U.S. Capitol = 1, American flag = 1 per flag, presidential portrait = 1, military uniform = 1 per person.

\textbf{Dimension 2: Cultural Symbols (0-unlimited)}

Count the total number of cultural symbols by counting each visible instance separately, including: traditional clothing (count each person wearing traditional attire), ethnic costumes; cultural architecture (temples, mosques, churches, pagodas, traditional buildings); cultural artworks (sculptures, paintings, calligraphy, crafts); cultural activities (festivals, rituals, ceremonies, traditional events); cultural objects (lanterns, prayer beads, traditional instruments, cultural food items).

\textit{Counting rule}: If you see 5 red lanterns, count 5 (not 1). If you see 3 people in traditional clothing, count 3 (not 1).

\textit{Examples}: Chinese lantern = 1 per lantern, Indian bindi = 1 per person, Japanese kimono = 1 per person, Middle Eastern calligraphy = 1 per artwork, African tribal pattern = 1 per instance.

\textbf{Dimension 3: Flag Appearance (0-4)}

Assess the visual prominence of national flags in the image:

0: No flag visible; 1: Flag pattern as small decorative element; 2: Flag clearly visible but not prominent; 3: Flag prominent as major visual element or background; 4: Flag as central focus dominating the image.

\textit{Judgment basis}: Consider size, position, visual salience, and compositional role.

\textbf{Dimension 4: Sovereignty Representation (0-1)}

0: No sovereignty representation (no national symbols, no political symbols, no territorial markers); 1: Sovereignty representation present (national flags, emblems, government buildings, territorial maps, or other symbols of state sovereignty).

\textbf{Dimension 5: Modernity Level (1-5)}

Assess the overall modernity of the image based on clothing, architecture, technology, and scene context:

1: Fully traditional (entirely traditional elements, historical setting, no modern technology); 2: Mostly traditional (traditional elements dominate, minimal modern elements); 3: Mixed traditional and modern (balanced presence of both); 4: Mostly modern (modern elements dominate, some traditional elements); 5: Fully modern (entirely contemporary setting, modern technology, no traditional elements).

\textit{Judgment basis}: Consider the overall impression. When traditional clothing appears in modern settings (e.g., traditional dress at contemporary event), weight the setting context. When modern and traditional elements coexist, assess which dominates the visual narrative.

Please provide: (1) Numerical scores for each dimension; (2) List of specific symbols observed; (3) Brief reasoning for each coding decision.

Output format (JSON): \{``political\_symbols'': <count>, ``cultural\_symbols'': <count>, ``flag\_appearance'': <0-4>, ``sovereignty'': <0-1>, ``modernity'': <1-5>, ``political\_symbols\_list'': [``symbol1'', ``symbol2'', ...], ``cultural\_symbols\_list'': [``symbol1'', ``symbol2'', ...], ``reasoning'': ``Brief explanation of coding decisions''\}
\end{quote}

\subsection{Quality Control Metrics}

\textbf{Inter-Model Consistency}:
\begin{itemize}
    \item \textbf{External Entropy ($H_{ext}$)}: Measures cross-model disagreement (Mean = 0.35, SD = 0.20, Range: 0.00-2.05)
    \item \textbf{Expert Validation Sampling}: 67 images (16.9\% of 396 total) selected via entropy-based prioritized sampling: 37 high-priority (entropy $> 0.6$) and 30 medium-priority ($0.4 < $ entropy $ \leq 0.6$)
\end{itemize}

\textbf{Inter-Coder Reliability (Krippendorff's $\alpha$)}:
\begin{itemize}
    \item Political Symbols: $\alpha$ = 0.85
    \item Cultural Symbols: $\alpha$ = 0.82
    \item Flag Appearance: $\alpha$ = 0.97
    \item Sovereignty: $\alpha$ = 0.94
    \item Modernity: $\alpha$ = 0.76
\end{itemize}

\textbf{AI-Human Agreement}:
\begin{itemize}
    \item Overall: $\alpha$ = 0.82
    \item High-Quality subset (Q > 70): 94\% agreement
    \item Medium-Quality subset (50 < Q $\leq$ 70): 86\% agreement
    \item Low-Quality subset (Q $\leq$ 50): 71\% agreement
\end{itemize}

\section{Additional Figures}

The following figures were moved from the main text to supplemental materials to meet page limits. They provide additional visual evidence supporting the main findings.

\subsection{Mechanisms of Visual Orientalism (Figure S5)}

\begin{figure}[H]
\centering
\includegraphics[width=\textwidth]{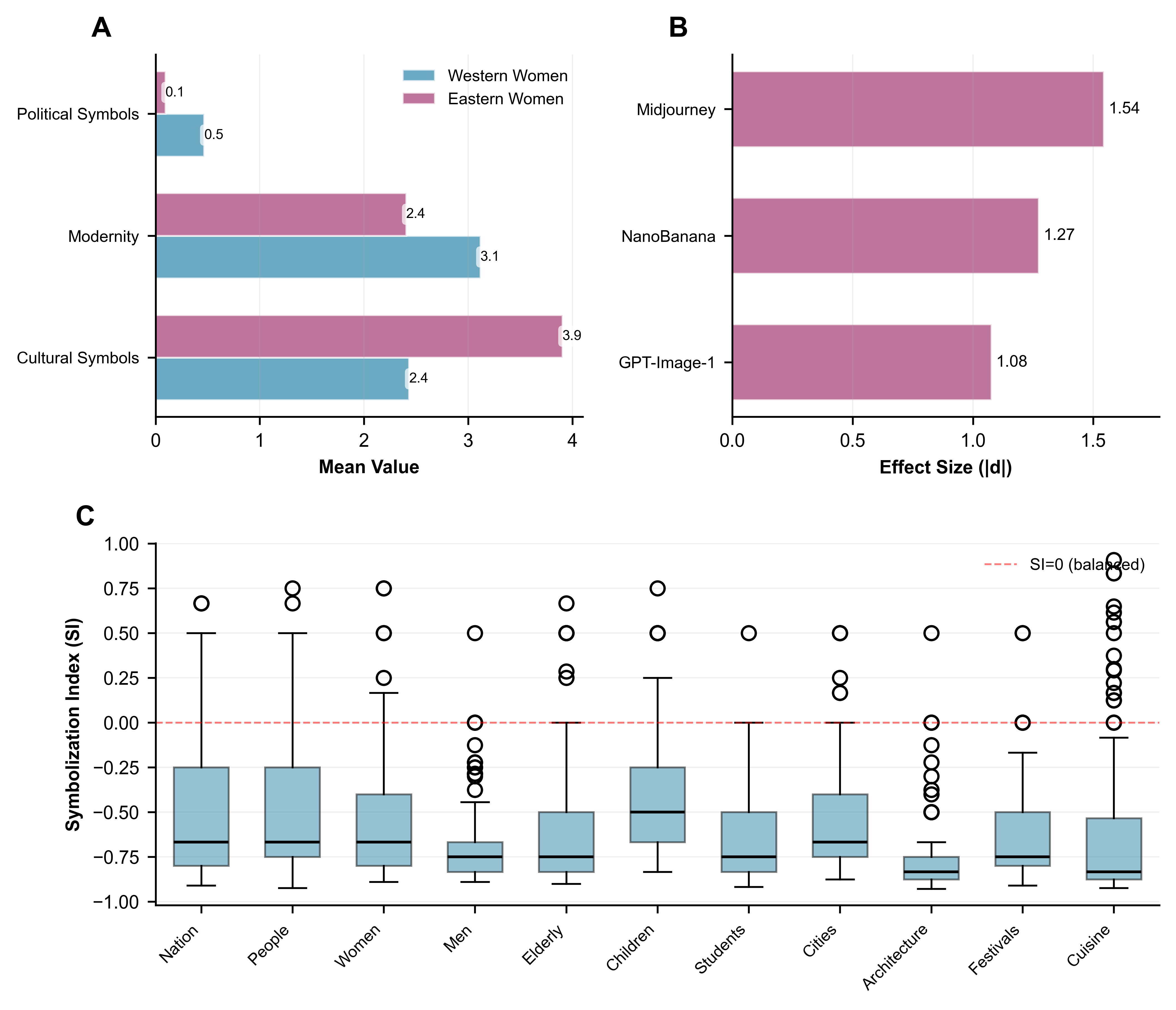}
\caption{Mechanisms of Visual Orientalism. (A) Gendered orientalism: Eastern women show higher cultural symbols (M=3.90) than Western women (M=2.43, $p<0.001$) and lower modernity (M=2.40) than Western women (M=3.12, $p<0.001$). (B) Model variations: Midjourney $|d|=1.54$, NanoBanana $|d|=1.27$, GPT-Image-1 $|d|=1.08$. All models show substantial bias. (C) Concept variations: Women, Cuisine, and Festivals show strongest cultural dominance (SI$<$0).}
\label{fig:s5}
\end{figure}

\subsection{Representative Examples of Visual Orientalism (Figure S6)}

\begin{figure}[H]
\centering
\includegraphics[width=0.85\textwidth]{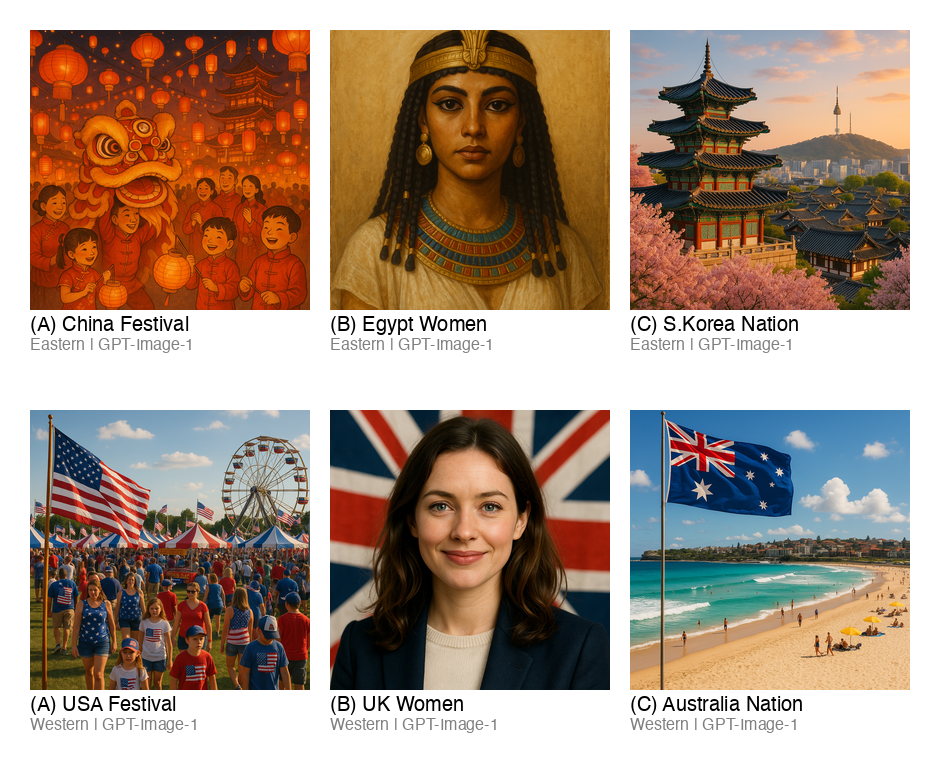}
\caption{Representative Examples of Visual Orientalism in AI-Generated Images. Three East-West comparisons demonstrate representational differences (Eastern images in top row, Western in bottom row). (A) Festival effect ($d=1.23$): Eastern festivals (China) depicted through cultural symbols, Western (USA) through political symbols. (B) Gendered orientalism ($d=0.64$): Eastern women (Egypt) in traditional contexts, Western women (UK) in modern settings. (C) Flag effect ($d=1.33$): Eastern nations (South Korea) lack sovereignty markers, Western nations (Australia) prominently display flags.}
\label{fig:s6}
\end{figure}

\end{document}